\documentclass[letterpaper,twoside,english]{emulateapj}
\usepackage[T1]{fontenc}
\usepackage{color}
\usepackage{booktabs}
\usepackage{graphicx}

\makeatletter

\providecommand{\tabularnewline}{\\}

\usepackage{apjfonts}
\shorttitle{Polarization in NGC 891}
\shortauthors{SEON}

\makeatother

\usepackage{babel}
\begin{document}

\title{Polarization as a Probe of Thick Dust Disk in Edge-on Galaxies: \\Application
to NGC 891}

\author{Kwang-il Seon\altaffilmark{1,2}}

\altaffiltext{1}{Korea Astronomy and Space Science Institute, Daejeon, 34055, Korea; kiseon@kasi.re.kr}
\altaffiltext{2}{Astronomy and Space Science Major, Korea University of Science and Technology, Daejeon, 34113, Korea}
\begin{abstract}
Radiative transfer models were developed to understand the optical
polarizations in edge-on galaxies, which are observed to occur even
outside the geometrically thin dust disk, with a scale height of $\approx0.2$
kpc. In order to reproduce the vertically extended polarization structure,
we find it is essential to include a geometrically thick dust layer
in the radiative transfer model, in addition to the commonly-known
thin dust layer. The models include polarizations due to both dust
scattering and dichroic extinction which is responsible for the observed
interstellar polarization in the Milky Way. We also find that the
polarization level is enhanced if the clumpiness of the interstellar
medium, and the dichroic extinction by vertical magnetic fields in
the outer regions of the dust lane are included in the radiative transfer
model. The predicted degree of polarization outside the dust lane
was found to be consistent with that (ranging from 1\% to 4\%) observed
in NGC 891.
\end{abstract}


\section{INTRODUCTION}

It is well known that materials in late-type spiral galaxies are mostly
concentrated at the galactic midplane. However, there have been many
attempts to probe extraplanar dust existing outside the galactic midplane.
High-resolution optical images have revealed filamentary dust complexes
above the galactic midplane up to $\sim2$ kpc in nearby edge-on spiral
galaxies \citep{1997AJ....114.2463H,1999AJ....117.2077H,2000AJ....119..644H,2000A&A...356..795A,2004AJ....128..674R,2004AJ....128..662T}.
The filamentary structures are traced by absorption against the background
starlight, thereby implying preferentially ``dense'' dust clouds with
an optical depth of $\gtrsim1$. Therefore, a relatively ``diffuse''
dust component, if any, was not detectable in these studies. The ``diffuse''
dust should appear as a faint extended reflection nebula illuminated
by starlight when the scale height of the stars is smaller than that
of the extraplanar dust. The ultraviolet (UV) reflection halo due
to the extraplanar dust residing above the galactic plane was first
discovered in NGC 891 \citep{2012IAUS..284..135S,2014ApJ...785L..18S}.
\citet{2014ApJ...789..131H} and \citet{2016ApJ...833...58H} reported
detection of the UV halos around many highly-inclined galaxies. \citet{2014ApJ...785L..18S}
and \citet{2015ApJ...815..133S} developed radiative transfer models
for the UV maps of edge-on galaxies to estimate the spatial distribution
and amount of the extraplanar dust.

Polarization maps in the visible and near-infrared (NIR) wavelengths
also allow large-scale galactic dust distributions to be traced \citep{1990IAUS..140..245S,1996MNRAS.278..519S,1996A&A...308..713F,1997AJ....114.1393J,2014ApJ...786...41M}.
The polarization in the optical wavelengths arises from dust scattering
of starlight and by dichroic extinction. Dichroic extinction is the
selective attenuation of starlight as it passes through a media in
which aspherical dust gains are aligned. Radiative torque and other
processes tend to align the long axes of aspherical dust grains in
a direction perpendicular to a magnetic field \citep[e.g.,][]{1996ASPC...97..401R,2007MNRAS.378..910L}.
The extinction cross section is larger along the longer axes of grains
so that light is preferentially absorbed along that direction. This
yields a net polarization parallel to the magnetic field direction.
On the other hand, the scattered light is polarized perpendicular
to the scattering plane, which gives a circular polarization pattern
around the central source. In any case, the presence of dust grains
is a necessary condition for polarization in the optical wavelengths.

In edge-on galaxies, for instance NGC 891 and NGC 4565, extended optical
polarization features were found in the bulge/halo regions above the
galactic midplane \citep{1990IAUS..140..245S,1996MNRAS.278..519S,1996A&A...308..713F}.
If only a thin dust layer exists in the galactic plane, polarization
arising from scattering and/or dichroic extinction is not expected
to be detected at high altitudes. It is, therefore, interesting to
investigate whether the extended optical polarization
is caused by the extraplanar dust layer which is responsible for the
UV reflection halo.

In this paper, Monte Carlo radiative transfer calculations are presented
to investigate the polarization pattern observed in an edge-on galaxy,
NGC 891. Large-scale geometry of the magnetic field in NGC 891 is
also discussed to explain the observed polarization patterns.

\section{RADIATIVE TRANSFER MODEL}

In order to explain the extended polarization pattern observed above
the dust lanes of edge-on galaxies, in particular, NGC 891, radiative
transfer models of dust-scattered and direct starlight were calculated
using the three-dimensional Monte Carlo radiative transfer code MoCafe
\citep{2014ApJ...785L..18S,2015JKAS...48...57S,2016ApJ...833..201S}.
The code models multiple scatterings of photons and uses a scheme
that includes ``forced first scattering'' to improve the calculation
efficiency in optically thin medium, and a ``peeling-off'' technique
to produce an image toward the observer. The basic Monte-Carlo algorithms
have been described in detail by many authors \citep{1996ApJ...465..127B,2001ApJ...551..269G,2011ApJS..196...22B,2013ARA&A..51...63S}.
The code was updated to take the polarization effect by dust scattering
into account, using an algorithm similar to that described in \citet{1996ApJ...465..127B}
and \citet{2017A&A...601A..92P}. Benchmark tests were performed as
described in \citet{2017A&A...601A..92P} and results similar to theirs
were obtained. The code was also verified using the polarization models
given in \citet{1996ApJ...465..127B}.

The dust grains are assumed to be spherical and to have a size distribution
as given by the MRN model \citep{1977ApJ...217..425M}: $n(a)\propto a^{-3.5}$
($0.005\mu{\rm m}<a<0.25\mu{\rm m}$), where $a$ is the grain radius.
The dielectric constants for the two dust compositions (astronomical
silicates and graphites) are adopted from the ones given by \citet{1984ApJ...285...89D}
and updated by \citet{2003ApJ...598.1026D}. All the relevant properties
(absorption and scattering cross sections, albedo, and Mueller matrix)
were calculated using the Mie scattering theory. We used the numerical
phase functions that were calculated with the Mie theory, instead
of using the approximate Henyey-Greenstein function. Polycyclic aromatic
hydrocarbons are not included because they are not relevant for the
present purpose.

The dichroic extinction is calculated according to a prescription
given by \citet{1997ApJ...477L..25W} and \citet{1997AJ....114.1405W}.
\citet{1989ApJ...346..728J} found that the magnitude of dichroic
polarization at the K band ($P_{K}$) is roughly proportional to $2.23\tau_{K}^{3/4}(\%)$
over a wide range of optical depths $\tau_{K}$. \citet{2008ApJ...674..304W}
also found a similar relation between the polarization and optical
depth: $P_{K}/\tau_{K}\approx5.08\tau_{V}^{-0.52}$ (\%). \citet{1997ApJ...477L..25W}
and \citet{1997AJ....114.1405W} assumed that the proportionality
relation of \citet{1989ApJ...346..728J} holds over other wavelengths
and obtained the polarization degree of $P_{V}=1.3\tau_{V}^{3/4}$
(\%) at the V band as a function of optical depth by combing the Serkowski
law and the Milky Way interstellar extinction curve, as given in \citet{1989AJ.....98.2062J}.
We also assume that the polarization relation measured at K holds
at V, but adopted the relation found by \citet{2008ApJ...674..304W},
and derived the polarization degree of $P_{V}=2.7\tau_{V}^{0.48}$
(\%) as a function of optical depth. At each interaction point, the
photon packet is polarized by an amount of $P_{V}\sin^{2}\theta$
in a direction parallel to the local magnetic field. Here, $\theta$
is the angle between the photon propagation direction and the magnetic
field direction at the interaction location. The photon packet is
also polarized by an amount calculated through the Mueller matrix
as it is scattered. The difference caused by the adopted dust models
does not significantly alter the present results and will be discussed
later.

The galaxy model is composed of two stellar (disk + bulge) and two
dust (thin and thick disks) components. The stellar and dust disks
are described by ``double-exponential'' distributions
\begin{equation}
\rho^{{\rm disk}}(r,z)=\rho_{0}^{{\rm disk}}\exp\left(-r/R-\left|z\right|/Z\right),\label{eq:1}
\end{equation}
where $\rho$ is the star or dust density at the radial and vertical
coordinates $r$ and $z$ in the cylindrical coordinate system. $\rho_{0}$
is the density at the galactic center. The radial scale length and
vertical scale height are denoted by $R$ and $Z$, respectively.
The luminosity density for the spheroidal bulge is given by the de-projection
function of a Sersic profile. The density profile with a profile index
$n$ is given by

\begin{equation}
\rho^{{\rm bulge}}(r,z)=\rho_{0}^{{\rm bulge}}\exp\left(-b_{n}B^{1/n}\right)B^{-(2n-1)/2n},\label{eq:2}
\end{equation}
where
\begin{equation}
B=\frac{\sqrt{r^{2}+z^{2}/q^{2}}}{R_{e}}.\label{eq:3}
\end{equation}
Here, $q$ is the minor to major axis ratio and $R_{e}$ the effective
radius. In this study, a de Vaucouleurs' profile corresponding to
a Sersic profile with $n=4$ ($b_{n}=7.67$) was used.

For the stellar and dust distributions, three types of models were
investigated, as shown in Table \ref{table1}. In the table, the optical
depths of the dust disks are the central face-on optical depths of
the exponential disks. Model A includes only a thin dust disk while
Models B, C and D include an additional, geometrically
thick dust disk. The model parameters for Model A are adopted from
the best-fit values estimated for the V-band data of NGC 891 in \citet{1999A&A...344..868X}.
The parameters for the thick dust disk, used to represent the extraplanar
dust in Models B, C and D, were adopted from the
best-fit model for the Far-UV observation as given by \citet{2014ApJ...785L..18S}.
Model B is the same as Model A except that the bulge effective radius
and the bulge-to-total luminosity ratio are slightly reduced compared
to those of Model A. The bulge light scattered by the extraplanar
dust disk makes the bulge appear bigger than it really is. Therefore,
the bulge parameters of Model B were slightly reduced in order to
produce a similar extent of the model galaxy as that obtained in Model
B, as will be shown later.

We also calculated radiative transfer models (Models
C and D) in a clumpy two-phase medium which consists of high-density
clumps and a low-density inter-clump medium. In Models
C and D, a ``smooth'' distribution of dust density $\left\langle \rho\right\rangle $
which is the same as that of Model B, was first produced. The probability
of any cell being in a high-density state or a low-density state is
randomly determined, based on the filling factor of the clumps ($f=0.2$
and 0.15 for Models C and D, respectively). The high-density clumps
are multiplied by a factor of $\rho_{{\rm high}}/\left\langle \rho\right\rangle =4$
(Model C) and 6 (Model D), and the low-density medium
by a factor of $\rho_{{\rm low}}/\left\langle \rho\right\rangle =(1-f\rho_{{\rm high}}/\left\langle \rho\right\rangle )/(1-f)=0.25$
(Model C) and 0.12 (Model D). The resulting medium is statistically
the same as that of Model B. The medium for Model
D is clumpier than that for Model C.

In general, edge-on galaxies show optical polarization within the
dust lane with orientations parallel to the galactic plane \citep{1990IAUS..140..245S,1996QJRAS..37..297S,1996A&A...308..713F,1997AJ....114.1393J,2014ApJ...786...41M}.
On the other hand, polarization in the outer regions appears to be
perpendicular to the galactic plane. These observational results imply
that there is a large-scale toroidal magnetic field in the inner regions
of galaxies and a vertical magnetic field in the outer regions. We,
therefore, considered three cases with respect to the magnetic field
geometry for each model type in Table \ref{table1}. In the first
case, no dichroic effect was included, to investigate the polarization
effect due to pure scattering. In the second case, the magnetic field
geometry was assumed to be toroidal, i.e., parallel to the galactic
plane and perpendicular to the radial direction in the cylindrical
coordinates. In the third case, the magnetic field was perpendicular
to the galactic plane, to mimic a poloidal magnetic field geometry.
The two magnetic field geometries are rather simple, but appear to
be good enough to demonstrate the dependence of polarization pattern
on the magnetic field geometry.

\section{RESULTS}

\begin{figure}
\begin{centering}
\medskip{}
\par\end{centering}
\begin{centering}
\includegraphics[clip,scale=0.8]{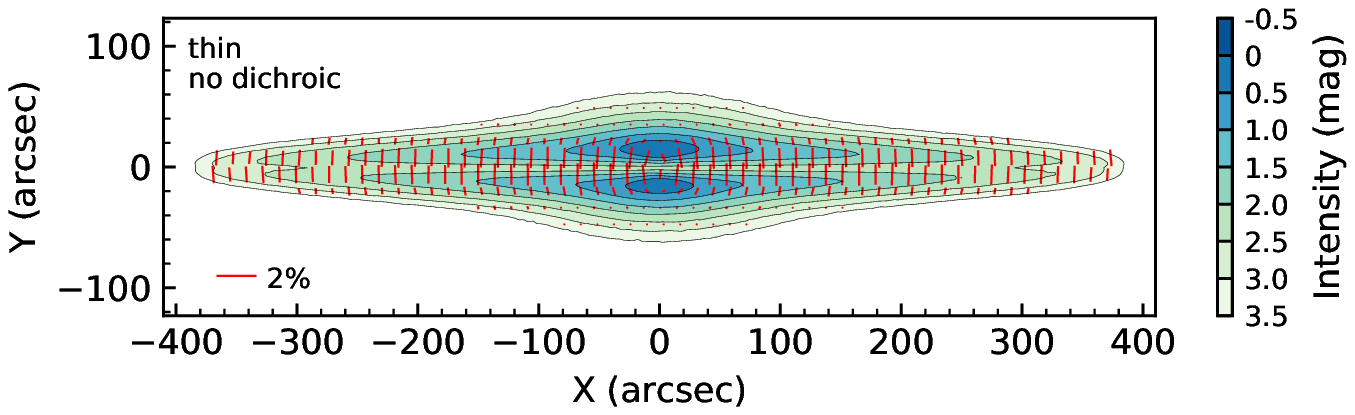}
\par\end{centering}
\begin{centering}
\includegraphics[clip,scale=0.8]{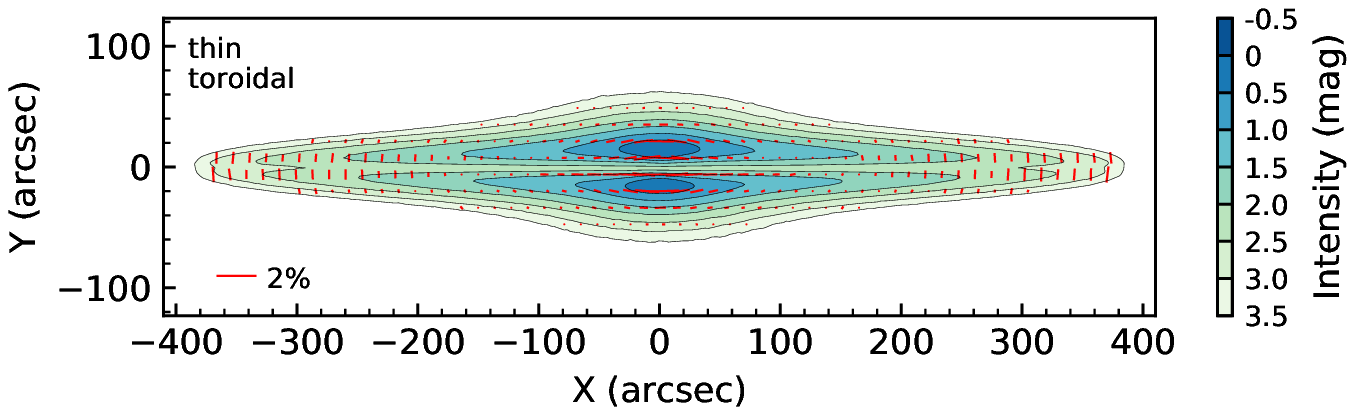}
\par\end{centering}
\begin{centering}
\includegraphics[clip,scale=0.8]{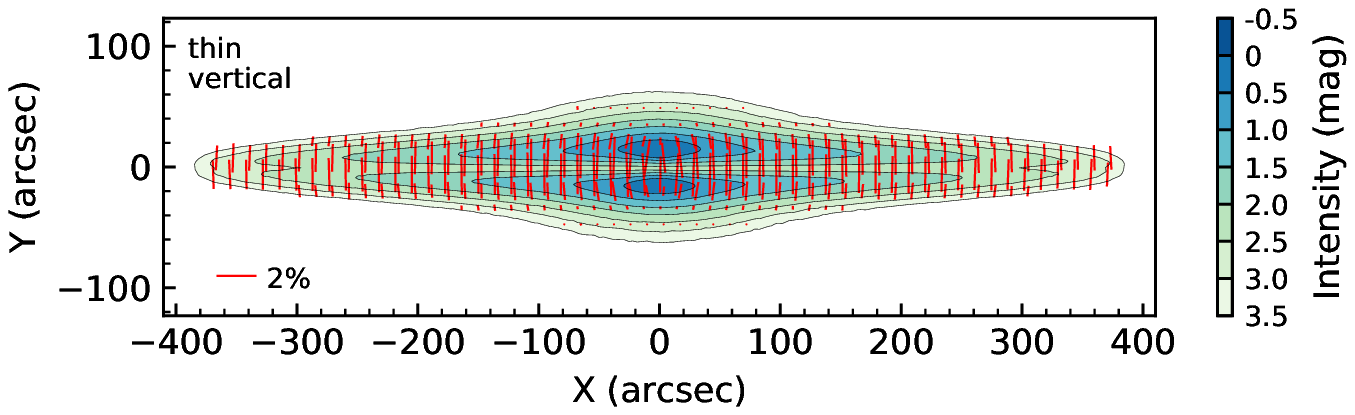}
\par\end{centering}
\begin{centering}
\medskip{}
\par\end{centering}
\caption{\label{fig1}V-band polarization maps for Model A in which only a
thin dust disk is considered. (top) No dichroic extinction is included.
(middle) Magnetic field is assumed to be toroidal (parallel to the
galactic plane). (bottom) Magnetic field is assumed to be vertical
(perpendicular to the plane). The intervals between contours are in
units of half-magnitude.}

\centering{}\medskip{}
\end{figure}

Figure \ref{fig1} shows the simulated polarization maps at V for
Model A, in which only a thin dust disk is included in the radiative
transfer calculation. In the top panel, no dichroic extinction is
included in the calculation. The polarization is found to be mostly
vertical to the galactic plane (centrosymmetric around the bulge)
and restricted within about $\pm20\arcsec$ ($\pm1.0$ kpc at the
distance of NGC 891, 9.5 Mpc) from the plane. Degree of polarization
is less than $\sim1$\%. The highest polarizations of $\sim1$\% occur
within the dust lane ($\left|Y\right|\lesssim5\arcsec$ or 0.25 kpc)
from the plane. We also note that the polarization is weaker in the
bulge region than in the outer regions. These are consistent with
the previous results obtained for edge-on galaxy models in \citet{1996ApJ...465..127B}
and \citet{2017A&A...601A..92P}.

If the magnetic field is parallel to the galactic plane (middle panel),
the polarization in the inner regions is very small and parallel to
the galactic plane. This is attributed to dichroic extinction by dust
grains aligned by toroidal magnetic fields. On the other hand, vertical
polarizations are shown in the outer regions in which scattering is
dominant. There are polarization null points near a radius of $100\arcsec$
in the dust lane. In the null points, the vertical polarization due
to scattering is cancelled out by the parallel polarization due to
the dichroic extinction along the toroidal magnetic field. The null
points are indeed found in optical polarization maps of edge-on galaxies
\citep{1990IAUS..140..245S,1995MNRAS.277.1430D,2014ApJ...786...41M}.

If the magnetic field is vertical to the plane (bottom panel), the
dichroic polarization enhances the polarization degree due to scattering.
However, in all three cases in Figure \ref{fig1}, the polarization
is very small or negligible outside the dust lane. This is because
the stellar scale height is much larger than the dust scale height.
At high galactic altitudes, stars are viewed through a very small
amount of dust and hence the polarization arising from scattering
or dichroic extinction is very low.

The above models are, however, in contrast to the NIR and visible
polarization maps, which show significant polarization structures
outside the dust lane \citep{1996MNRAS.278..519S,1996QJRAS..37..297S,1996A&A...308..713F,1997AJ....114.1393J}.
In particular, a degree of polarization of up to 4\% was found in
the bulge/halo regions of NGC 891 by \citet{1996A&A...308..713F}.
The polarization patterns observed in the bulge/halo regions suggest
the existence of a large scale dust layer above the midplane.

Figure \ref{fig2} shows the results for the case in which an additional,
thick dust disk with a scale height of 1.6 kpc and a central face-on
optical depth of 0.3 at V is also included. It can be immediately
recognized that the polarization signature is now visible even at
the bulge and high altitudes in all three cases. In the middle panel,
the null polarization points occur at about $100\arcsec$ in the dust
lane, as in the middle panel of Figure \ref{fig1}. The degree of
polarization is about 1.2\textendash 1.5 times higher than the case
in which no thick dust disk is included. In the case of vertical magnetic
field (bottom panel), up to $\sim3$\% polarization degree is found,
which is close to the maximum polarization of $\sim4$\% measured
in NGC 891 \citep{1996A&A...308..713F}. The polarization vectors
within $\sim50\arcsec$ from the galactic center are mainly perpendicular
to the plane. It is also clear that the bottom panel with the vertical
magnetic field shows higher degree of polarizations compared to the
two other cases, in which the magnetic field is absent or parallel
to the galactic plane. We also note that the degree of polarization
in general rises with distance from the midplane, as observed in the
optical polarization maps.

The galaxy models in Figures \ref{fig1} and \ref{fig2} consist of
smooth dust and stellar components. For better models, we need to
consider the clumpiness of the interstellar medium (ISM). The polarization
map for Model C, in which the ISM is clumpy, is shown in Figure \ref{fig3}.
A polarization degree of up to $\sim4$\% is found, which is consistent
with the observational result of \citet{1996A&A...308..713F}. Figure
\ref{fig4} shows the polarization map for Model D. The resulting
degree of polarization in both models is higher than in Figure \ref{fig2}.
It is also found that a clumpier model (Model D) always yields higher
polarization than a less clumpy model (Model C) under the same magnetic
field structure. The degree of polarization appears to be equal to
or greater than 4\% only when both the dichroic extinction and clumpiness
are taken into account. We also note that the overall extent of the
galaxy map is larger than those in Figures \ref{fig1} and \ref{fig2}.
This is because the clumpier medium results in less efficient absorption
and more extended scattering features compared to the smoother medium.
We obtained similar polarization patterns even when the model parameters
were adjusted to make the resulting surface brightness map resemble
that shown in Figures \ref{fig1} and \ref{fig2}.

\section{DISCUSSION}

\label{sec:4}

In this study, we developed radiative transfer models which reproduce
the overall structure of the optical polarization maps observed in
NGC 891. We found that the extended optical polarization provides
strong evidence for the existence of the extraplanar dust which has
been inferred from the observations of ultraviolet reflection halos.

There is further evidence suggesting the existence of a thick dust
disk. \citet{2014ApJ...789..131H} examined the possibility that the
UV halos in highly-inclined galaxies originate from stars in the halos.
However, they found that the dust scattering nebula model is most
consistent with the UV observations. Observations in the mid-IR and
far-IR (FIR) wavelengths also provide evidence of the extraplanar
dust \citep{2000A&A...356..795A,2006A&A...445..123I,2007ApJ...668..918B,2013A&A...556A..54V,2016A&A...586A...8B}.
In particular, \citet{2016A&A...586A...8B} found that the FIR emission
in NGC 891 is best fit with the sum of a thin and a thick dust component,
and the scale height of the thick dust component was consistent with
the result given by \citet{2014ApJ...785L..18S}.

\citet{2014ApJ...785L..18S} found that the UV halos in NGC 891 were
well reproduced by a radiative transfer model with two exponential
dust disks, one with a scale height of $\approx$ 0.2\textendash 0.25
kpc and the other with a scale height of $\approx$ 1.2\textendash 2.0
kpc. The central face-on optical depth of the geometrically thick
disk was found to be $\approx$ 0.3\textendash 0.5 at the B band ($\approx$
0.23\textendash 0.38 at V). In this study, we demonstrated that the
extended polarization structures can be well reproduced by thick dust
disk with a scale height of 1.6 kpc and a central face-on optical
depth of 1.3 at V. We also varied the scale height and optical depth
of the thick dust disk within the ranges found in \citet{2014ApJ...785L..18S}.
The degree of polarization was slightly lowered as the optical depth
or the scale height was altered. However, no significant difference
was found.

\begin{figure}[t]
\begin{centering}
\medskip{}
\par\end{centering}
\begin{centering}
\includegraphics[clip,scale=0.8]{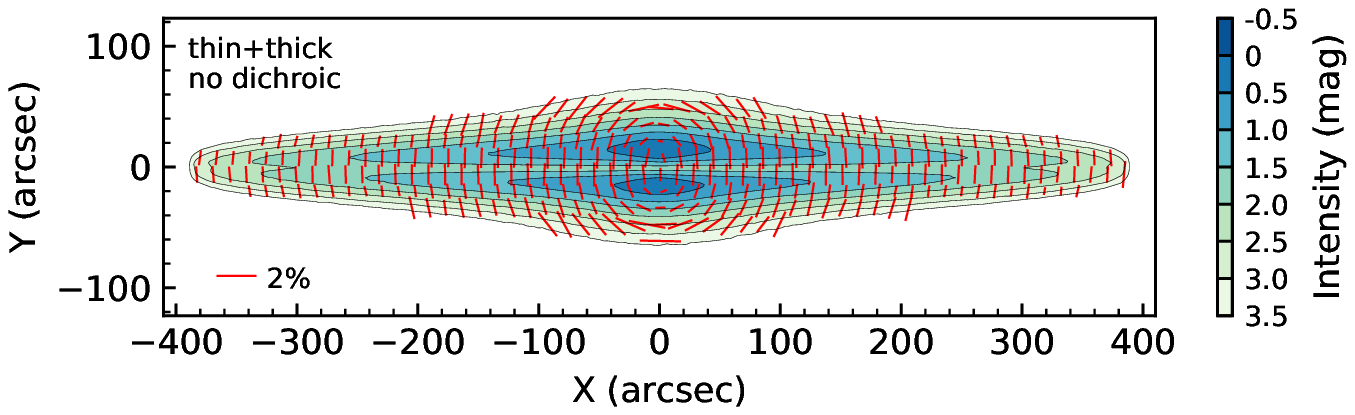}
\par\end{centering}
\begin{centering}
\includegraphics[clip,scale=0.8]{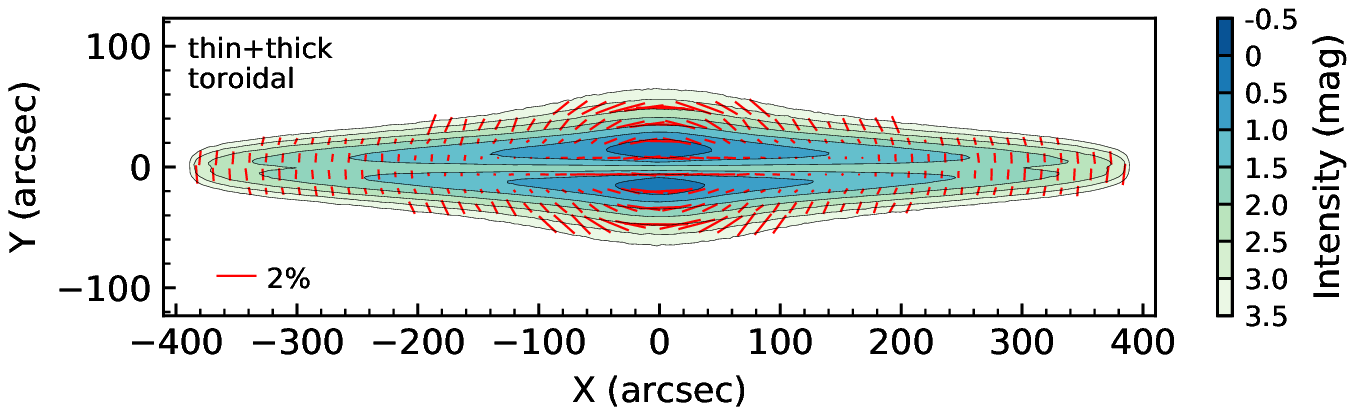}
\par\end{centering}
\begin{centering}
\includegraphics[clip,scale=0.8]{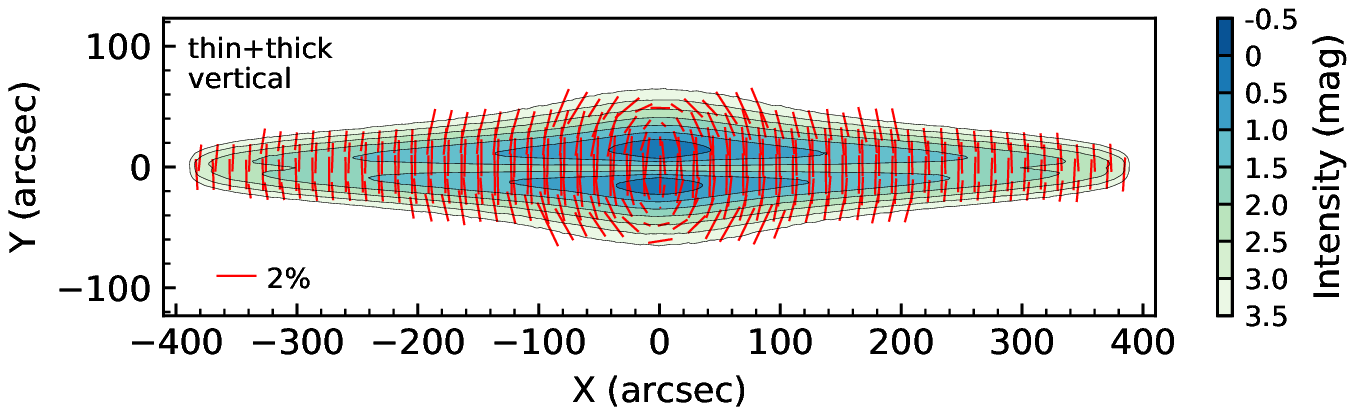}
\par\end{centering}
\begin{centering}
\medskip{}
\par\end{centering}
\caption{\label{fig2}V-band polarization maps for Model B in which both thin
and thick dust disks are considered. (top) No dichroic extinction
is included. (middle) Magnetic field is assumed to be toroidal (parallel
to the galactic plane). (bottom) Magnetic field is assumed to be vertical
(perpendicular to the plane).}

\centering{}\medskip{}
\end{figure}

\begin{figure}
\begin{centering}
\medskip{}
\par\end{centering}
\begin{centering}
\includegraphics[clip,scale=0.8]{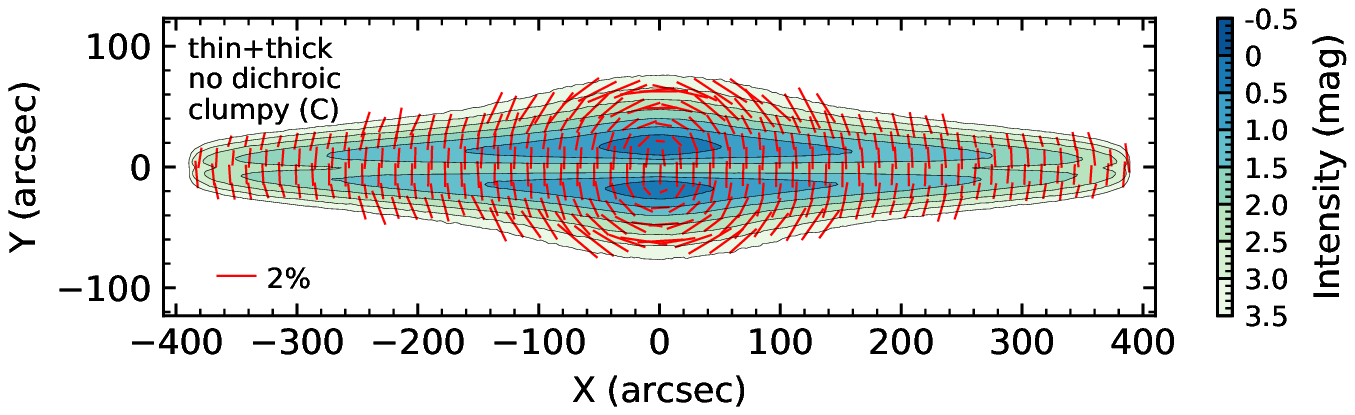}
\par\end{centering}
\begin{centering}
\includegraphics[clip,scale=0.8]{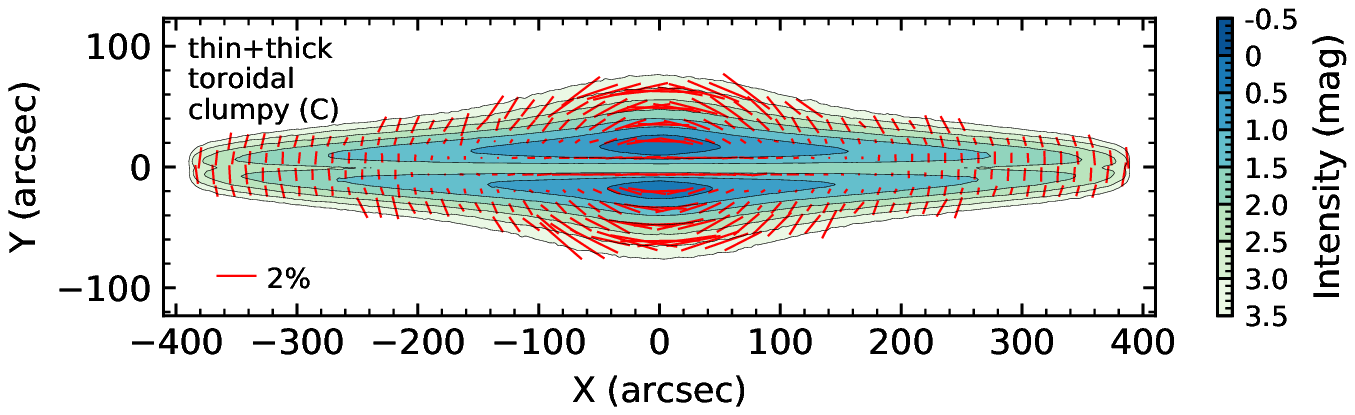}
\par\end{centering}
\begin{centering}
\includegraphics[clip,scale=0.8]{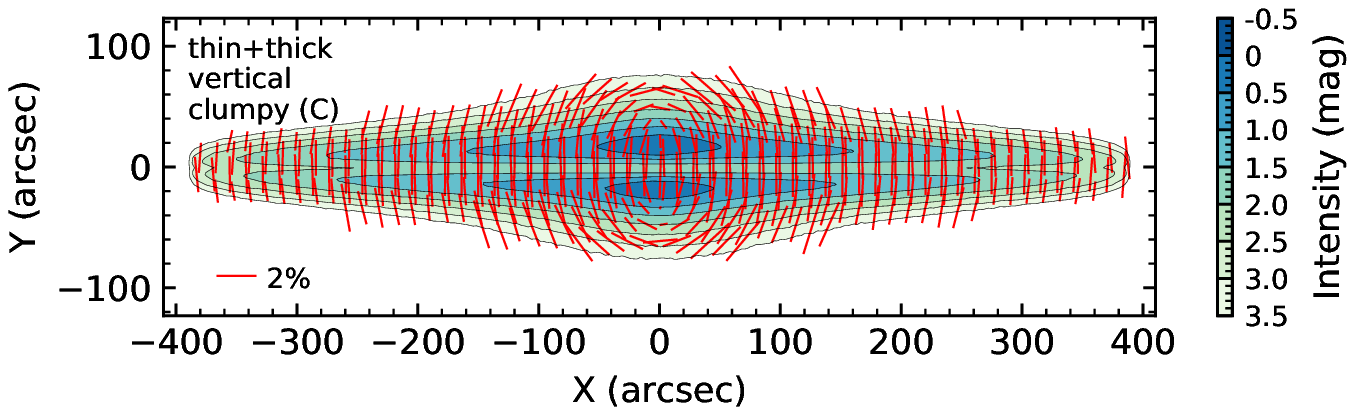}
\par\end{centering}
\begin{centering}
\medskip{}
\par\end{centering}
\caption{\label{fig3}V-band polarization maps for Model C in which the medium
is assumed to be clumpy. Magnetic field is assumed to be vertical.
(top) No dichroic extinction is included. (middle) Magnetic field
is assumed to be toroidal (parallel to the galactic plane). (bottom)
Magnetic field is assumed to be vertical (perpendicular to the plane).}

\centering{}\medskip{}
\end{figure}

\begin{figure}
\begin{centering}
\medskip{}
\par\end{centering}
\begin{centering}
\includegraphics[clip,scale=0.8]{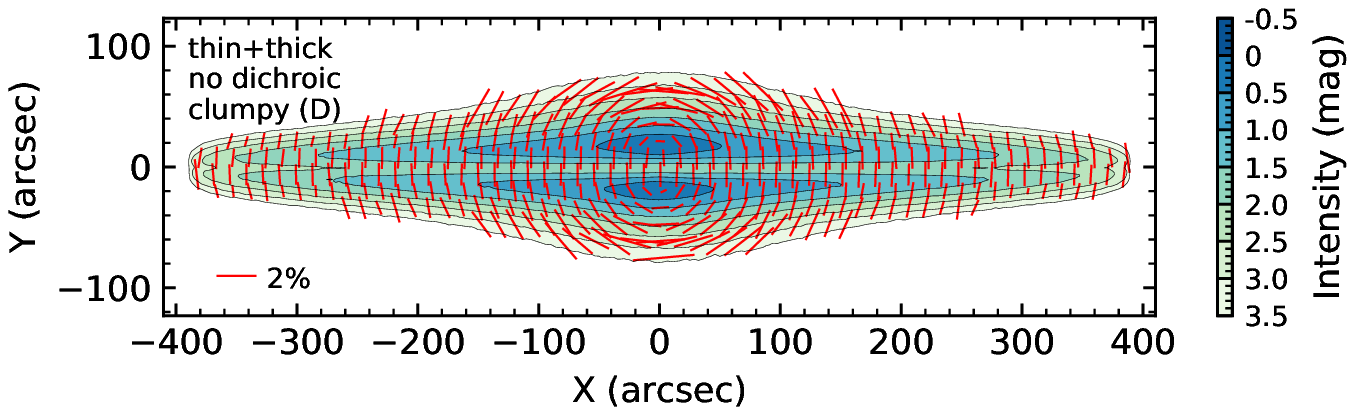}
\par\end{centering}
\begin{centering}
\includegraphics[clip,scale=0.8]{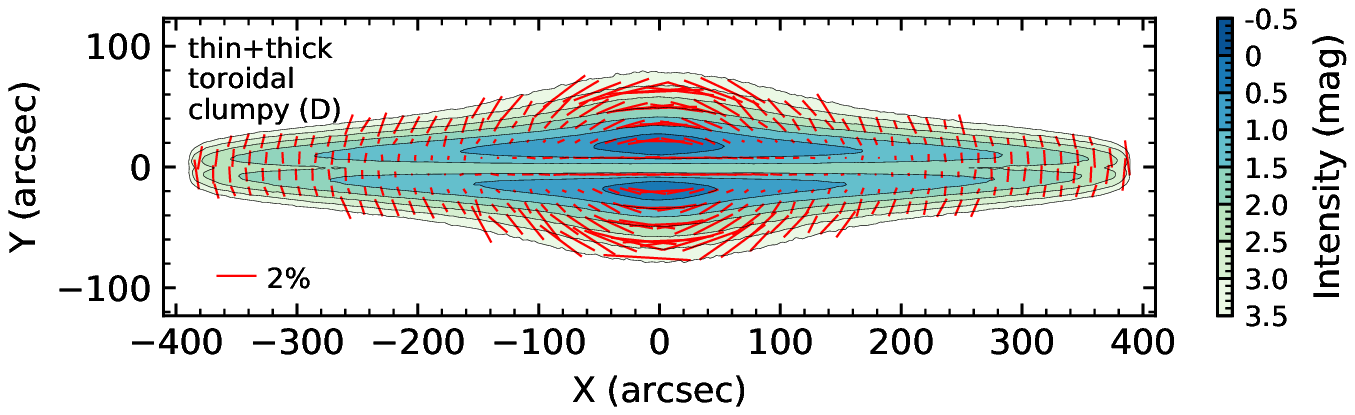}
\par\end{centering}
\begin{centering}
\includegraphics[clip,scale=0.8]{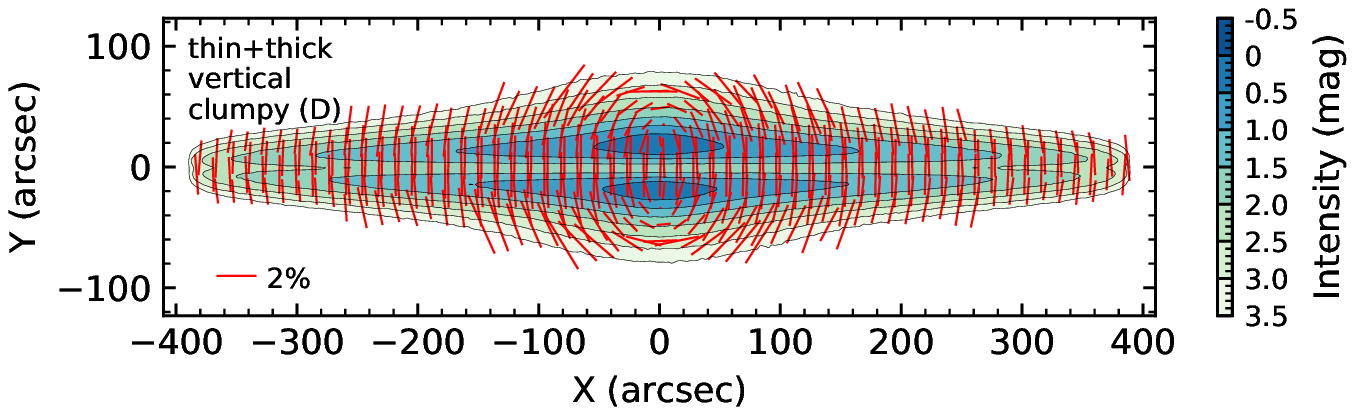}
\par\end{centering}
\begin{centering}
\medskip{}
\par\end{centering}
\caption{\label{fig4}V-band polarization maps for Model D in which the medium
is assumed to be clumpy. Magnetic field is assumed to be vertical.
The medium for Model D is clumpier than that for Model C. (top) No
dichroic extinction is included. (middle) Magnetic field is assumed
to be toroidal (parallel to the galactic plane). (bottom) Magnetic
field is assumed to be vertical (perpendicular to the plane).}

\centering{}\medskip{}
\end{figure}

The observed polarization vectors, especially at V-band,
are in general parallel to the galactic plane in the central part,
but perpendicular to the galactic plane in the outer regions. Therefore,
we may conclude that the magnetic fields are in general
vertical (or poloidal) except the central part, where the magnetic
fields are mainly toroidal. This is consistent with the result obtained
from polarized radio emission \citep{1991ApJ...382..100S,1995A&A...302..691D,2009RMxAC..36...25K}.
It is clear that the vertical polarizations in the outer regions can
mostly be attributed to scattered starlight. However, without the
effect of dichroic extinction and/or clumpiness, the
predicted degree of polarization is slightly lower than the observed
values (middle panel of Figure \ref{fig2}). In our
models, both the dichroic polarization and the clumpiness are also
required to reproduce the observed polarization of up to 4\% \citep{1996A&A...308..713F}.

However, it is noteworthy that \citet{2014ApJ...786...41M}
found a different V- and H-band orientation pattern in the outer disk
of NGC 891. They argue that polarization verctors are unreliable tracers
of the magnetic field. The optical polarization vector orientation
on the northeast side of the galaxy appears to differ between \citet{1996MNRAS.278..519S}
and \citet{1996A&A...308..713F}. Therefore, a definite conclusion
on the magnetic field orientation cannot be drawn from the V-band
data alone. We also note that the polarization vector within the
dust lane will be very low or erratic if a large-scale magnetic field
is absent or the magnetic fields are more or less random in the central
region. This would be the case of NGC 891, in which no clear pattern
of polarization vectors are found in the dust lane \citep{1996MNRAS.278..519S,1996A&A...308..713F}.

The maximum degree of polarization for a single scattering in our
dust model is only marginally lower than that calculated in \citet{1979ApJ...229..954W},
but significantly greater than the model of \citet{2001ApJ...548..296W}.
For instance, the maximum polarization degree at the V band is 36\%
in our dust model, while 38\% and 24\% were found in the models of
\citet{1979ApJ...229..954W} and \citet{2001ApJ...548..296W}, respectively.
The dust model of \citet{1979ApJ...229..954W} used the same MRN's
size distribution of dust grains, but adopted different dielectric
constants than ours. The polarization degree due to dichroic extinction
assumed in \citet{1997AJ....114.1405W} is $P_{V}=1.3\tau_{V}^{3/4}$
(\%), which is lower than the relation ($P_{V}=2.7\tau_{V}^{0.48}$)
used in this study. It was found that the degree of dichroic polarization
was $P_{V}=2.2\tau_{V}^{3/4}$ (\%) if the polarization relation at
K obtained in \citet{1989ApJ...346..728J} and the MRN dust model
are combined in deriving the relation. The differences are mainly
caused by uncertainties in the polarization relation and the dust
extinction cross section. The overall polarization level was slightly
reduced when the relation $P_{V}=1.3\tau_{V}^{3/4}$ (\%) was adopted,
but no substantial difference was found.

Note that we assumed spherical dust grains when calculating the Mueller
matrix and adopted an empirical approach to take into account the
dichroic polarization effect. A more detailed treatment of the dichroic
polarization can be found in \citet{2014A&A...566A..65R} and \citet{2016A&A...593A..87R}.
In their approach, the scattering phase function depends not only
on the incident photon direction but also on the alignment of dust
grains. We also note that the clumpy models (Models
C and D) were devised to demonstrate that the clumpiness is also
an important factor in studying polarizations. Spiral
partterns were not included in the present radiative transfer models.
It may be necessary to use more realistic density models. However,
more detailed models are beyond the scope of the present work.

\acknowledgements{This work was supported by the National Research Foundation of Korea
(NRF) grant funded by the Korea government (MSIP) (No. 2017R1A2B4008291).
Numerical simulations were partially performed by using a high performance
computing cluster at the Korea Astronomy and Space Science Institute.}

\newpage{}

\begin{table}[tp]
\caption{\label{table1}Model parameters}

\begin{centering}
\begin{tabular}{llr@{\extracolsep{0pt}.}lr@{\extracolsep{0pt}.}l}
\toprule 
Parameter &  & \multicolumn{2}{c}{Model A} & \multicolumn{2}{c}{Model B, C, D}\tabularnewline
\midrule
Scale height of stellar disk & $Z_{s}$ & 0&4 & 0&4\tabularnewline
Scale length of stellar disk & $R_{s}$ & 5&4 & 5&4\tabularnewline
Bulge to total luminosity ratio & B/T & 0&3 & 0&2\tabularnewline
Bulge effective radius & $R_{e}$ & 1&5 & 1&0\tabularnewline
Bulge axial ratio & $q$ & 0&5 & 0&5\tabularnewline
Optical depth of thin dust disk & $\tau_{V}^{{\rm thin}}$ & 0&8 & 0&8\tabularnewline
Scale height of thin dust disk & $Z_{d}^{{\rm thin}}$ & 0&25 & 0&25\tabularnewline
Scale length of thin dust disk & $R_{d}^{{\rm thin}}$ & 7&7 & 7&7\tabularnewline
Optical depth of thick dust disk & $\tau_{V}^{{\rm thick}}$ & \multicolumn{2}{c}{} & 0&3\tabularnewline
Scale height of thick dust disk & $Z_{d}^{{\rm thick}}$ & \multicolumn{2}{c}{} & 1&6\tabularnewline
Scale length of thick dust disk & $R_{d}^{{\rm thick}}$ & \multicolumn{2}{c}{} & 7&7\tabularnewline
Inclination angle & $\theta_{{\rm inc}}$ & 89&8 & 89&8\tabularnewline
\bottomrule
\end{tabular}
\par\end{centering}
\centering{}\medskip{}
\end{table}

\begin{thebibliography}{}
\expandafter\ifx\csname natexlab\endcsname\relax\def\natexlab#1{#1}\fi

\bibitem[{Alton {et~al.}(2000)Alton, Xilouris, Bianchi, Davies, \&
  Kylafis}]{2000A&A...356..795A}
Alton, P.~B., Xilouris, E.~M., Bianchi, S., Davies, J., \& Kylafis, N. 2000,
  A{\&}A, 356, 795

\bibitem[{Baes {et~al.}(2011)Baes, Verstappen, De~Looze, Fritz, Saftly,
  Vidal~P{\'e}rez, Stalevski, \& Valcke}]{2011ApJS..196...22B}
Baes, M., Verstappen, J., De~Looze, I., {et~al.} 2011, ApJS, 196, 22

\bibitem[{Bianchi {et~al.}(1996)Bianchi, Ferrara, \&
  Giovanardi}]{1996ApJ...465..127B}
Bianchi, S., Ferrara, A., \& Giovanardi, C. 1996, ApJ, 465, 127

\bibitem[{Bocchio {et~al.}(2016)Bocchio, Bianchi, Hunt, \&
  Schneider}]{2016A&A...586A...8B}
Bocchio, M., Bianchi, S., Hunt, L.~K., \& Schneider, R. 2016, A{\&}A, 586, A8

\bibitem[{Burgdorf {et~al.}(2007)Burgdorf, Ashby, \&
  Williams}]{2007ApJ...668..918B}
Burgdorf, M., Ashby, M. L.~N., \& Williams, R. 2007, ApJ, 668, 918

\bibitem[{Draine(2003)}]{2003ApJ...598.1026D}
Draine, B.~T. 2003, ApJ, 598, 1026

\bibitem[{Draine \& Lee(1984)}]{1984ApJ...285...89D}
Draine, B.~T., \& Lee, H.~M. 1984, ApJ, 285, 89

\bibitem[{Draper {et~al.}(1995)Draper, Done, Scarrott, \&
  Stockdale}]{1995MNRAS.277.1430D}
Draper, P.~W., Done, C., Scarrott, S.~M., \& Stockdale, D.~P. 1995, MNRAS, 277,
  1430

\bibitem[{Dumke {et~al.}(1995)Dumke, Krause, Wielebinski, \&
  Klein}]{1995A&A...302..691D}
Dumke, M., Krause, M., Wielebinski, R., \& Klein, U. 1995, A{\&}A, 302, 691

\bibitem[{Fendt {et~al.}(1996)Fendt, Beck, Lesch, \&
  Neininger}]{1996A&A...308..713F}
Fendt, C., Beck, R., Lesch, H., \& Neininger, N. 1996, A{\&}A, 308, 713

\bibitem[{Gordon {et~al.}(2001)Gordon, Misselt, Witt, \&
  Clayton}]{2001ApJ...551..269G}
Gordon, K.~D., Misselt, K.~A., Witt, A.~N., \& Clayton, G.~C. 2001, ApJ, 551,
  269

\bibitem[{Hodges-Kluck \& Bregman(2014)}]{2014ApJ...789..131H}
Hodges-Kluck, E., \& Bregman, J.~N. 2014, ApJ, 789, 131

\bibitem[{Hodges-Kluck {et~al.}(2016)Hodges-Kluck, Cafmeyer, \&
  Bregman}]{2016ApJ...833...58H}
Hodges-Kluck, E., Cafmeyer, J., \& Bregman, J.~N. 2016, ApJ, 833, 58

\bibitem[{Howk \& Savage(1997)}]{1997AJ....114.2463H}
Howk, J.~C., \& Savage, B.~D. 1997, AJ, 114, 2463

\bibitem[{Howk \& Savage(1999)}]{1999AJ....117.2077H}
---. 1999, AJ, 117, 2077

\bibitem[{Howk \& Savage(2000)}]{2000AJ....119..644H}
---. 2000, AJ, 119, 644

\bibitem[{Irwin \& Madden(2006)}]{2006A&A...445..123I}
Irwin, J.~A., \& Madden, S.~C. 2006, A{\&}A, 445, 123

\bibitem[{Jones(1989{\natexlab{a}})}]{1989ApJ...346..728J}
Jones, T.~J. 1989{\natexlab{a}}, ApJ, 346, 728

\bibitem[{Jones(1989{\natexlab{b}})}]{1989AJ.....98.2062J}
---. 1989{\natexlab{b}}, AJ, 98, 2062

\bibitem[{Jones(1997)}]{1997AJ....114.1393J}
---. 1997, AJ, 114, 1393

\bibitem[{Krause(2009)}]{2009RMxAC..36...25K}
Krause, M. 2009, RMxAC, 36, 25

\bibitem[{Lazarian \& Hoang(2007)}]{2007MNRAS.378..910L}
Lazarian, A., \& Hoang, T. 2007, MNRAS, 378, 910

\bibitem[{Mathis {et~al.}(1977)Mathis, Rumpl, \&
  Nordsieck}]{1977ApJ...217..425M}
Mathis, J.~S., Rumpl, W., \& Nordsieck, K.~H. 1977, ApJ, 217, 425

\bibitem[{Montgomery \& Clemens(2014)}]{2014ApJ...786...41M}
Montgomery, J.~D., \& Clemens, D.~P. 2014, ApJ, 786, 41

\bibitem[{Peest {et~al.}(2017)Peest, Camps, Stalevski, Baes, \&
  Siebenmorgen}]{2017A&A...601A..92P}
Peest, C., Camps, P., Stalevski, M., Baes, M., \& Siebenmorgen, R. 2017,
  A{\&}A, 601, A92

\bibitem[{Reissl {et~al.}(2016)Reissl, Wolf, \& Brauer}]{2016A&A...593A..87R}
Reissl, S., Wolf, S., \& Brauer, R. 2016, A{\&}A, 593, A87

\bibitem[{Reissl {et~al.}(2014)Reissl, Wolf, \& Seifried}]{2014A&A...566A..65R}
Reissl, S., Wolf, S., \& Seifried, D. 2014, A{\&}A, 566, A65

\bibitem[{Roberge(1996)}]{1996ASPC...97..401R}
Roberge, W.~G. 1996, in Polarimetry of the interstellar medium. Astronomical
  Society of the Pacific Conference Series; Vol. 97; Proceedings of a
  conference held at Rensselaer Polytechnic Institute; Troy; New York; 4-7 June
  1995; San Francisco: Astronomical Society of the Pacific (ASP); |c1996;
  edited by Wayne G. Roberge and Doug C. B. Whittet, 401

\bibitem[{Rossa {et~al.}(2004)Rossa, Dettmar, Walterbos, \&
  Norman}]{2004AJ....128..674R}
Rossa, J., Dettmar, R.-J., Walterbos, R. A.~M., \& Norman, C.~A. 2004, AJ, 128,
  674

\bibitem[{Scarrott(1996)}]{1996QJRAS..37..297S}
Scarrott, S.~M. 1996, QJRAS, 37, 297

\bibitem[{Scarrott \& Draper(1996)}]{1996MNRAS.278..519S}
Scarrott, S.~M., \& Draper, P.~W. 1996, MNRAS, 278, 519

\bibitem[{Scarrott {et~al.}(1990)Scarrott, Rolph, \&
  Semple}]{1990IAUS..140..245S}
Scarrott, S.~M., Rolph, C.~D., \& Semple, D.~P. 1990, in IN: Galactic and
  intergalactic magnetic fields; Proceedings of the 140th Symposium of IAU,
  Durham, University, England, 245--251

\bibitem[{Seon(2015)}]{2015JKAS...48...57S}
Seon, K.-I. 2015, JKAS, 48, 57

\bibitem[{Seon \& Draine(2016)}]{2016ApJ...833..201S}
Seon, K.-I., \& Draine, B.~T. 2016, ApJ, 833, 201

\bibitem[{Seon \& Witt(2012)}]{2012IAUS..284..135S}
Seon, K.-I., \& Witt, A.~N. 2012, The Spectral Energy Distribution of Galaxies,
  284, 135

\bibitem[{Seon {et~al.}(2014)Seon, Witt, Shinn, \& Kim}]{2014ApJ...785L..18S}
Seon, K.-I., Witt, A.~N., Shinn, J.-H., \& Kim, I.-J. 2014, ApJ, 785, L18

\bibitem[{Shinn \& Seon(2015)}]{2015ApJ...815..133S}
Shinn, J.-H., \& Seon, K.-I. 2015, ApJ, 815, 133

\bibitem[{Steinacker {et~al.}(2013)Steinacker, Baes, \&
  Gordon}]{2013ARA&A..51...63S}
Steinacker, J., Baes, M., \& Gordon, K.~D. 2013, ARA{\&}A, 51, 63

\bibitem[{Sukumar \& Allen(1991)}]{1991ApJ...382..100S}
Sukumar, S., \& Allen, R.~J. 1991, ApJ, 382, 100

\bibitem[{Thompson {et~al.}(2004)Thompson, Howk, \&
  Savage}]{2004AJ....128..662T}
Thompson, T. W.~J., Howk, J.~C., \& Savage, B.~D. 2004, AJ, 128, 662

\bibitem[{Verstappen {et~al.}(2013)Verstappen, Fritz, Baes, Smith, Allaert,
  Bianchi, Blommaert, De~Geyter, de~Looze, Gentile, Gordon, Holwerda, Viaene,
  \& Xilouris}]{2013A&A...556A..54V}
Verstappen, J., Fritz, J., Baes, M., {et~al.} 2013, A{\&}A, 556, 54

\bibitem[{Weingartner \& Draine(2001)}]{2001ApJ...548..296W}
Weingartner, J.~C., \& Draine, B.~T. 2001, ApJ, 548, 296

\bibitem[{White(1979)}]{1979ApJ...229..954W}
White, R.~L. 1979, ApJ, 229, 954

\bibitem[{Whittet {et~al.}(2008)Whittet, Hough, Lazarian, \&
  Hoang}]{2008ApJ...674..304W}
Whittet, D. C.~B., Hough, J.~H., Lazarian, A., \& Hoang, T. 2008, ApJ, 674, 304

\bibitem[{Wood(1997)}]{1997ApJ...477L..25W}
Wood, K. 1997, ApJL, 477, L25

\bibitem[{Wood \& Jones(1997)}]{1997AJ....114.1405W}
Wood, K., \& Jones, T.~J. 1997, AJ, 114, 1405

\bibitem[{Xilouris {et~al.}(1999)Xilouris, Byun, Kylafis, Paleologou, \&
  Papamastorakis}]{1999A&A...344..868X}
Xilouris, E.~M., Byun, Y.~I., Kylafis, N.~D., Paleologou, E.~V., \&
  Papamastorakis, J. 1999, A{\&}A, 344, 868

\end{thebibliography}
\end{document}